\documentclass[final,3p,times,twocolumn,hdvipdfm]{elsarticle}

\usepackage{amssymb}

\usepackage{amsmath}
\usepackage[all, warning]{onlyamsmath}
\usepackage[breaklinks=true]{hyperref} 

\journal{Physics Letters B}

\usepackage[normalem]{ulem} 
\usepackage{color}

\newcommand{\cSM}[1]{{\color{black}#1}}

\begin{document}


\begin{frontmatter}



\title{
A Walking Dilaton Inflation} 


\author[KEK]{Hiroyuki~Ishida}

\ead{ishidah@post.kek.jp}

\author[JLU]{Shinya~Matsuzaki\corref{cor}}

\ead{synya@jlu.edu.cn}

\cortext[cor]{Corresponding author}

\address[KEK]{Theory Center, IPNS, KEK, Tsukuba, Ibaraki 305-0801, Japan}
\address[JLU]{Center for Theoretical Physics and College of Physics, Jilin University, Changchun, 130012,
China}

\begin{abstract}
We propose an inflationary scenario based on a many-flavor hidden QCD with eight flavors, 
which realizes  
the almost scale-invariant (walking) gauge dynamics. 
The theory predicts two types of composite (pseudo) Nambu-Goldstone bosons,  
the pions and the lightest scalar (dilaton) 
associated with the spontaneous chiral symmetry breaking and its simultaneous 
violation of the approximate scale invariance. 
The dilaton acts as an inflaton, where the inflaton potential is induced by 
the nonperturbative-scale anomaly linked with the underlying theory. 
The inflaton potential parameters 
are highly constrained by the walking nature, which are evaluated 
by straightforward nonperturbative analyses including lattice simulations. 
Due to the pseudo Nambu-Goldstone boson's natures and the intrinsic  
property for the chiral symmetry breaking in the walking gauge dynamics,  
the inflaton coupled to the pions 
naturally undergoes the small field inflation consistently 
with all the cosmological and astrophysical constraints presently placed by 
Planck 2018 data. 
When the theory is vector-likely coupled to the standard model in part in a way to realize 
a dynamical electroweak symmetry breaking, 
the reheating temperature is determined by the pion decays to electroweak gauge 
bosons. The proposed inflationary scenario would provide a dynamical origin 
for the small field inflation as well as the light pions as a smoking-gun to be probed 
by future experiments.

\end{abstract}


\end{frontmatter}


\section{Introduction}

Exponentially expanded cosmological evolution, inflation, 
is one of the most attractive and plausible period of our Universe 
to dynamically and simultaneously solve crucial cosmological problems, 
such as the flatness, the homogeneous and isotropic, and the horizon problems. 
The dynamics itself would have no doubt that we have experienced such an epoch, 
however, details of inflation have not been revealed 
so far both theoretically and experimentally at all.

As one of candidate scenarios for the inflation, 
models having a $\log$-type potential can be good targets. 
Since the $\log$-type potential naturally has a plateau, 
it can be easily applied to the inflation. 
Such a $\log$-type potential can be given in the context of 
the Coleman-Weinberg (CW) mechanism~\cite{Coleman:1973jx} 
where quantum loop contributions modify the shape of the potential to be flatten around at the origin, \cSM{which is called the} CW inflation~\cite{Barenboim:2013wra}. 
However, it is in general quite difficult to satisfy all the cosmological parameters 
fitted by cosmic microwave background (CMB) observations 
such as the Planck satellite~\cite{Aghanim:2018eyx}. 
Some recent developments have been made where 
the small field inflation (SFI) of the CW type can be achieved 
without conflicting with \cSM{the CMB} observations 
by introducing a linear term originated to a fermion condensation~\cite{Iso:2014gka} 
and a dynamical origin of the initial field value~\cite{Iso:2015wsf}. 
Still, it is however unavoidable 
to naively assume the quartic coupling of the inflaton 
to be extremely small, in order to 
realize the observed amplitude of the scalar perturbation.

At this moment, one should notice that 
the inflaton quartic coupling $\lambda_\chi$ can be expressed by 
the ratio of the inflaton mass $(m_\chi)$ to the vacuum expectation value of 
the inflaton $(v_\chi)$, $\lambda_\chi \sim (m_\chi/v_\chi)^2$, with 
the stationary condition taken into account. 
This implies that the tiny enough $\lambda_\chi$ is directly 
linked to a large enough scale hierarchy between $m_\chi$ and $v_\chi$. 
Then, an interesting question would be raised: 
{\it ``Can this large hierarchy be physical in a sense of quantum field theory? ''}

\cSM{It is a walking gauge theory that} can naturally supply 
such a large scale hierarchy, which is characterized by 
the ``journey-distance'' during the walking (i.e. almost scale-invariant) 
behavior, from 
an ultraviolet (UV) scale $(\Lambda_{\rm UV})$ 
down to an infrared (IR) scale $(\Lambda_{\rm IR})$.

For instance, in the case of many flavor QCD with the 
SU(N) group coupled to fermions $(F)$ belonging to 
the fundamental representation, 
the walking regime can be established 
reflecting the perturbative IR fixed point, well known as the 
Caswell-Banks-Zaks IR fixed point~\cite{Caswell:1974gg,Banks:1981nn}. 
In that case, the fermion dynamical mass scale $(m_F)$ 
is expected to be generated from the UV scale $\Lambda_{\rm UV}$ 
having the characteristic scaling relation, called Miransky scaling~\cite{Miransky:1984ef} 
(sometimes also called Berezinsky-Kosterlitz-Thouless scaling) 
intrinsic to the conformal phase transition~\cite{Miransky:1996pd}: 
$m_F \sim \Lambda_{\rm UV} e^{-\pi/\sqrt{\alpha/\alpha_c -1}}$ 
(in the chiral broken phase), 
where $\alpha$ is the fine structure constant of the gauge theory 
and $\alpha_c$ denotes the critical coupling above which the 
chiral symmetry breaking takes place. 
This scaling surely realizes a large scale hierarchy in the 
chiral broken phase during the walking regime, 
spanned by the IR $m_F$ and the UV $\Lambda_{\rm UV}$ scales, 
where the $\Lambda_{\rm UV}$ can be identified as 
the critical scale at which the $m_F$ is generated.

The lightest composite scalar, so-called walking dilaton, 
arises as the consequence of the spontaneous breaking 
of the (approximate) scale invariance, and simultaneously 
gets massive due to the explicit breaking induced by  
the scale anomaly~\cite{Leung:1985sn} arising from the Miransky scaling above: 
$\beta(\alpha) = \partial \alpha/\partial \ln \Lambda_{\rm UV}  
\sim  - \frac{\alpha_c}{\pi} (\alpha/\alpha_c -1)^{3/2}$. 
QCD with eight flavors has been confirmed 
by lattice simulations   
to be walking with the chiral broken phase~\cite{Aoki:2013xza,Appelquist:2014zsa,Hasenfratz:2014rna}. 
 In that case, 
 it has been observed on lattices~\cite{Aoki:2014oha,Aoki:2016wnc,Appelquist:2018yqe} 
that the walking dilaton formed by a flavor singlet bilinear $\bar{F}F$ 
can be as light as or less than the chiral symmetry breaking scale $m_F$, 
in accordance with the expected particle identity as a 
pseudo Nambu-Goldstone (pNG) boson for the scale symmetry breaking. 
Furthermore, it has been found~\cite{Aoki:2014oha,Aoki:2016wnc} that 
the dilaton decay constant ($f_\chi$) can be much larger than 
the $m_F$, in accord with a many-flavor version of the Veniziano limit, 
dubbed anti-Veneziano limit~\cite{Matsuzaki:2015sya}:  
$f_\chi \sim \sqrt{N N_F} m_F$ for $N \alpha=$fixed, 
$N_F/N=$fixed $ \gg 1$, and $N, N_F \to \infty$. 
Hence the $f_\chi$, dictated by the scale-chiral breaking, 
would be on the order of the chiral-critical scale $\Lambda_{\rm UV}$ above. 
Thus one would have a large scale hierarchy between 
$m_\chi \lesssim m_F$ and $f_\chi \sim \Lambda_{\rm UV}$, via 
the Miranksky scaling, $m_\chi \sim f_\chi e^{-\pi/\sqrt{\alpha/\alpha_c-1}}$, 
which has indeed been supported by some straightforward  
Schwinger-Dyson equation-analysis on many-flavor walking gauge 
theories~\cite{Hashimoto:2010nw}.

Note that the dilaton potential is also generated by the scale anomaly 
driven by the chiral symmetry breaking, 
and is fixed by the anomalous Ward-Takahashi identity for the scale symmetry~\cite{Matsuzaki:2015sya,Matsuzaki:2012vc,Matsuzaki:2013eva},    
to generically take the form of CW type~\footnote{
A similar composite dilaton (or glueball) potential of the CW type was discussed in 
a context different from the present interest in the SFI, where 
a large field inflation with non-minimal coupling to general relativity 
is assumed~\cite{Channuie:2011rq,Bezrukov:2011mv,Channuie:2013lla,Channuie:2013xoa,Channuie:2014kda,Channuie:2014ysa,Channuie:2016xmq}. }, like  
$V_\chi \sim \lambda_\chi \chi^4 \ln \chi$ with $\lambda_\chi \sim (m_\chi/v_\chi)^2$ 
and $v_\chi$ being identified as the dilaton decay constant $f_\chi$ above 
($v_\chi \equiv f_\chi$). 
Thus, one undoubtedly expects that, with the walking dilaton identified as an inflaton,  
a tiny enough quartic coupling, 
$\lambda_\chi \sim (m_\chi/v_\chi)^2 
\equiv (m_\chi/f_\chi)^2 \sim (m_\chi/\Lambda_{\rm UV})^2 \ll 1$, 
desired for the consistent inflation,  
would naturally and dynamically be generated by the walking gauge dynamics, 
when the IR and UV scales are associated with $m_\chi$ and $v_\chi$, respectively.

A possible inflationary history goes like: 
We start from the chiral-scale broken phase in the vacuum for a many flavor 
walking gauge theory, where we have the walking dilaton and its potential 
of a CW type. Allowing some coupling between the walking fermions and 
some scalar, e.g. the standard model (SM)-like Higgs, a preheating mechanism~\cite{Kofman:1997yn} 
would work to trap the walking dilaton around the origin of the potential -- 
(``approximate'') chiral-scale symmetric point -- 
due to an induced-finite matter-density effect (like a thermal plasma) 
triggered by parametric resonances, 
as proposed in~\cite{Iso:2015wsf}. 
As the temperature cools down, the walking 
dilaton is going slowly to roll the potential-down hill, 
i.e. undergoes the SFI due to the underlying walking nature, which 
governs the evolution of Universe at that time. 
After ending the inflation (almost in the same manner as 
in the CW inflation), the dilaton starts to drop down to the vacuum $\chi=v_\chi$, 
oscillate and will keep reheating until it decays. 
The reheating would work 
unless the created temperature gets higher than the critical temperature(s)  
for the chiral phase transition and/or deconfinement phase transition -- 
expected to be around the temperature $\sim m_F$ -- above which 
the walking dilaton gets disassociated to cease having the potential.

In this paper, we present \cSM{a dynamical inflationary scenario of the CW-SFI type}   
arising from eight-flavor QCD  
(a large $N_F$ walking gauge theory). 
The walking 
dilaton plays the role of an inflaton, 
where the inflaton potential parameters 
are highly constrained by the walking nature. 
We evaluate the potential parameters 
using outputs from 
straightforward nonperturbative analyses including lattice simulations. 
It is shown that 
with a tiny dilaton coupling to pions explicitly breaking the chiral-scale 
symmetry, 
the walking dilaton inflation resolves a well-known incompatibility 
intrinsic to the SFI of the CW type 
for realizing the desired e-folding number and the observed 
spectral index~\cite{Barenboim:2013wra,Takahashi:2013cxa}. 
This makes the proposal in~\cite{Iso:2014gka} explicitized 
by a concrete dynamics as the many-flavor walking gauge theory.

To be more realistic, 
the inflationary scenario is fitted 
with all the cosmological and astrophysical constraints presently placed by 
Planck 2018 data~\cite{Aghanim:2018eyx} and 
theoretical requirements on the walking dilaton and chiral pion physics.   
As a reference scenario, 
the hidden walking dynamics 
is vector-likely 
gauged in part by the electroweak (EW) charges and is coupled to a Higgs doublet in 
a scale-invariant way. 
Thus the EW symmetry breaking (EWSB) is triggered by the bosonic seesaw mechanism~\cite{Calmet:2002rf,Kim:2005qb,Haba:2005jq,Antipin:2014qva,Haba:2015lka,Haba:2015qbz,Ishida:2016ogu,Ishida:2016fbp,Haba:2017wwn,Haba:2017quk,Ishida:2017ehu,Ishida:2019gri}, 
as a high-scale dynamical scalegenesis, in which the gauge hierarchy problem 
is possibly absent. 
We find that the present inflationary scenario is highly constrained 
to give a stringent bound on the fermion dynamical mass 
scale ($m_F$) to be greater than $10^{11}$ GeV, and  
the reheating temperature ($T_R$) is then fixed 
by chiral pion decays to EW 
gauge bosons including photons to be $\gtrsim 10^2$ GeV, 
where the masses of walking dilaton and pions  
are also constrained by several theoretical and astrophysical limits  
to be $\gtrsim 10^8$ GeV and $\gtrsim 10^5$ GeV, respectively. 
Since $T_R \ll m_F$, this reheating thus works consistently with 
the deconfinement and or chiral phase transitions in the walking 
gauge theory, as noted above. 

Thus the presently proposed inflationary scenario of the CW-SFI type 
would provide the dynamical origin and explanation for the tiny inflaton coupling and 
somewhat light particles (the walking pions) 
as a smoking-gun to be probed by future experiments.


\section{Walking dilaton potential}

We begin by writing the dilaton potential induced from 
a generic many-flavor QCD, which takes the form including 
the CW type: 
\begin{align}
 V(\chi) = - \frac{C}{{2 N_F}} \chi^a  { {\rm tr}[U + U^\dag] }  
 + \frac{\lambda_\chi}{4} \chi^4 
 \left( \ln \frac{\chi}{v_\chi} + A \right) + V_0  
 \,.\label{pot-chi}
\end{align}
$V_0$ is the vacuum energy, which is determined by taking 
the normalization of the potential as $V_0(v_\chi)=0$.

The $C$ term in Eq.(\ref{pot-chi}) has come from the explicit-chiral 
breaking (flavor universal) mass term 
for the hidden QCD fermions, 
\begin{align} 
{\cal L}_m =- m_0 \sum_{i=1}^{N_F} \bar{F}_iF_i
\,, \label{m0}
\end{align}  
(with $m_0$ being real), by extracting  
the flavor-singlet component as 
\begin{align} 
\bar{F}_{R i}F_{L j} \approx 
\langle \bar{F}_{R i} F_{L i}  \rangle \cdot 
\left(\frac{\chi}{v_\chi} \right)^{a} \cdot U_{ij} 
\,, 
\end{align}  
(and its hermitian conjugate partner).  
{$U = e^{2i\pi/f_\pi}$ is the chiral field 
parametrized by the pion field $\pi = \pi^\alpha T^\alpha$ 
with $T^\alpha$ being generators of SU($N_F$) 
($\alpha=1,\cdots, N_F^2-1$)  
normalized as ${\rm tr}[T^\alpha T^\beta]=\delta^{\alpha\beta}/2$,   
and the pion decay constant $f_\pi$.  
The exponent $a$ for the $\chi$ controls the size of the 
overlap amplitude between 
the $\chi$ and the composite operator $\bar{F}F$ in the underlying theory, 
for which we will take $a=1$, so 
the $\chi$ is allowed to linearly couple to 
the $\bar{F}F$}~\footnote{
As discussed in the literature~\cite{Leung:1989hw,Matsuzaki:2013eva}, 
when the walking dilaton $\chi$ purely arises as the $\bar{F}F$-composite scalar, 
the power parameter $a$ would be identical to $(3-\gamma_m)$, 
the dynamical dimension of the $\bar{F}F$ operator with the anomalous mass 
dimension $\gamma_m$, which is fixed by the anomalous Ward-Takahashi identity 
for the scale symmetry. 
However, possible mixing with other flavor-singlet 
scalars, like a glueball or tetraquark states, might be present, so in that sense 
the parameter $a$ would generically be undermined until fully solving 
the mixing structure, that is beyond the scope of the current interest. 
In the present work, thus, we will take a conservative 
limit with $a=1$, 
so that the $\chi$ couples to the $\bar{F}F$ as if it were a conventional 
singlet-scalar component as seen in the linear sigma model:    
the chiral SU(N$_F$)$_{L} \times$ SU(N$_F$)$_{R}$ linear sigma-model field $M$ ($M^\dag$) is introduced as the effective local-operator description for the $\bar{F}_R F_L$ ($\bar{F}_L F_R$),  and  
 transforms as $M \to g_L \cdot M \cdot g_R^\dag$ under the chiral  symmetry with the transformation matrices $g_{L,R}$. 
 The current (universal) $F$-fermion 
 mass $m_0$ is therefore coupled to the $M$ ($M^\dag$) so as 
 to respect the original form in Eq.(\ref{m0}) in a chiral invariant way, like ${\rm tr}[{\cal M} M^\dag + M {\cal M}^\dag]$ with the mass-parameter spurion field ${\cal M}$ transforming in the same way as $M$, with the vacuum value $\langle {\cal M} \rangle = m_0 \cdot 1_{{\rm N}_F \times {\rm N}_F}$. 
In the chiral broken phase, 
the $M$ can generically be polar-decomposed by hermitian ($\tilde{M}$) and unitary $(\xi_{L,R})$ matrices as $M = \xi_L \cdot \tilde{M} \cdot \xi_R$ where $\xi_L^\dag \xi_R=U$. Supposing a low-energy limit where only the lightest scalar ($\chi$) survives among the $N_F$ scalars in $\tilde{M}$, 
one may write $M \approx \chi \cdot U$. Plugging this approximated expression into the above ${\cal M}^\dag M + {\rm h.c.}$ term, 
to get $\chi \cdot m_0 (U + U^\dag)$, the coupling form of 
which coincides with the 
conservative limit $a=1$. Thus, the choice $a=1$ is reasonable 
if the linear sigma model gives a good low-energy description 
for the underlying walking gauge theory in terms of 
the chiral-breaking structure. }  
Hence the parameter $C$ is expressed {in terms of the pion mass $m_\pi$ and the decay constant $f_\pi$} as 
\begin{align} 
 { C = N_F \frac{m_\pi^2 f_\pi^2}{2 v_\chi}}  
 \,, \label{C-eq}
\end{align} 
with the canonical form of the pion kinetic term being assumed. 
Note that the $f_\pi$ can be related 
with the fermion dynamical mass $m_F$ as 
\begin{align} 
f_\pi \simeq \sqrt{N} \cdot \frac{m_F}{2 \pi}
\,, \label{fpi}
\end{align}
which is based on a naive dimensional analysis 
regarding the definition of the pion decay constant~\footnote{
{The Pagels-Stokar formula~\cite{Pagels:1979hd} applying to 
the present walking gauge dynamics (with the nonrunning and ladder 
approximation taken) would yield~\cite{Matsuzaki:2015sya} 
$f_\pi \simeq \sqrt{N/(2 \pi^2)} m_F$, which is larger by about factor of $\sqrt{2}$. 
}}. 
Then the parameter $C$ in Eq.(\ref{C-eq}) can be evaluated as 
\begin{align} 
C \simeq \frac{N N_F  m_\pi^2 m_F^2}{ {8 \pi^2} v_\chi}
\,. \label{C-mpi}
\end{align}

The quartic coupling $\lambda_\chi$ in Eq.(\ref{pot-chi}) is set by the ratio of 
the dilaton mass $m_\chi$ 
to the dilaton decay constant $v_\chi$ as $\lambda_\chi = (m_\chi/v_\chi)^2$ 
(by taking into account the stationary condition)
 in the absence of the explicit chiral-scale breaking by the $C$ term (i.e. chiral limit). 
To this chiral-limit quantity, 
some straightforward computation of the scale anomaly in the many-flavor 
walking gauge theory (called ladder Schwinger-Dyson equation analysis), 
in combination with the partially-conserved dilatation-current (PCDC) 
relation, would give a constraint~\cite{Matsuzaki:2015sya}~\footnote{
The right hand side corresponds to the vacuum expectation value of the trace of (symmetric part of) energy momentum tensor, $\langle \theta_\mu^\mu  \rangle = 4 {\cal E}_{\rm vac}$, where the ${\cal E}_{\rm vac}$  denotes the vacuum energy in the walking gauge theory, 
which is dominated by the $F$-fermion loop contribution to the gluon 
condensate~\cite{Matsuzaki:2015sya}, hence it scales solely with the 
dynamical mass $m_F$, like ${\cal E}_{\rm vac} \propto N N_F m_F^4$. 
On the other hand, the definition of the dilaton decay constant $f_\chi(=v_\chi)$ gives at the dilaton-soft mass limit $p^2 = m_\chi^2 \to 0$,   
$\langle 0| \theta_\mu^\mu(0) | \phi(p) \rangle = - m_\chi^2 f_\chi $ 
(with $\phi = f_\chi \log (\chi/f_\chi)$), 
the left hand side of which can be evaluated by the ${\cal E}_{\rm vac}$ 
by using the PCDC: $\theta_\mu^\mu(x)=- m_\chi^2 f_\chi e^{- i xp} \phi(x)$ together with the standard reduction formula, and  
the Ward-Takahashi identity for the scale symmetry (the 
low-energy theorem), as $\langle 0| \theta_\mu^\mu(0) | \chi(p) \rangle = -  
4 d_{\theta_\mu^\mu} \cdot {\cal E}_{\rm vac}/f_\chi$ with 
the scale dimension of $\theta_\mu^\mu$, $d_{\theta_\mu^\mu}=4$. 
Thus one has $m_\chi^2 f_\chi^2 = -16 {\cal E}_{\rm vac}$. For 
more detailed evaluation, see the literature~\cite{Matsuzaki:2015sya} and 
references therein.  
}:  
\begin{align} 
 m_\chi^2 v_\chi^2 \simeq \frac{16 N N_F}{\pi^4} m_F^4
 \,. \label{PCDC}
\end{align}
Thereby the $\lambda_\chi$ may be evaluated as 
\begin{align} 
 \lambda_\chi \simeq 
 \frac{16 N N_F}{\pi^4} \left(\frac{m_F}{v_\chi} \right)^4 
 \,. \label{lambda-chi-mF}
\end{align}
The full walking dilaton mass $(M_\chi)$ is given 
(by evaluating $V''(v_\chi)=\partial^2 V(\chi)/\partial \chi^2|_{\chi=v_\chi}$)  
as the sum of the chiral limit value ($m_\chi$) and the correction from 
the $C$-term~\footnote{{This mass formula is precisely the same as 
the one (with a factor $(3-\gamma_m)(1+\gamma_m)$ taken to be 3) 
derived in the dilaton-chiral perturbation theory for the many-flavor 
walking gauge theory at the leading order 
of the derivative expansion~\cite{Leung:1989hw,Matsuzaki:2013eva}.}}: 
\begin{align} 
 M_\chi^2 &= m_\chi^2 + \frac{3 C}{v_\chi} 
 \notag \\ 
 & { \left( \simeq m_\chi^2 + 3N_F \frac{m_\pi^2 f_\pi^2}{2 v_\chi^2}\right)} 
 { \simeq m_\chi^2 + 3 N N_F \frac{m_\pi^2 m_F^2}{8 \pi^2 v_\chi^2}} 
 \,. \label{Mchi}
\end{align}

Through the stationary condition (evaluated at $\chi=v_\chi$), 
the parameter $A$ in Eq.(\ref{pot-chi}) is given as a function in terms of the $C$ and $\lambda_\chi$ 
to be $A = - \frac{1}{4} + \frac{C}{\lambda_\chi v_\chi^3}$~\footnote{
The second term in this stationary condition, which gives a scale invariant 
$\chi^4$ term proportional to the chiral explicit 
breaking $C$ parameter, plays the role of stabilization of the dilaton 
potential in the presence of the fermion current mass, as 
pointed out it is necessary to have in the literature~\cite{Leung:1989hw,Matsuzaki:2013eva}.  
}.

Thus, the walking dilaton potential in Eq.(\ref{pot-chi}) 
is highly constrained by nontrivial parameter correlations 
given by Eqs.(\ref{C-mpi}) and (\ref{lambda-chi-mF}). 
Then the potential $V(\chi)$ normalized to $v_\chi^4(\equiv f_\chi^4)$ is 
essentially controlled by small underlying-theory parameters, 
$m_\pi/v_\chi \ll 1$ and $m_F/v_\chi \ll 1$ 
(with $m_\pi \ll m_F$, to be consistent with the nonlinear realization 
for the chiral-scale symmetry in which we are currently working, as 
seen from the potential form in Eq.(\ref{pot-chi})).

\section{Small field inflation}

The slow roll parameters ($\eta$ and $\epsilon$), the e-folding number $(N)$ 
and the magnitude of the 
scalar perturbation $(\Delta_R^2)$ are respectively defined as  
\begin{align} 
 \eta & = M_{\rm pl}^2 \left( \frac{V^{\prime \prime}(\chi)}{V(\chi)} \right) 
\,, \notag \\ 
\epsilon & = \frac{M_{\rm pl}^2}{2} \left( \frac{V^{\prime}(\chi)}{V(\chi)} \right)^2  
\,, \notag \\ 
N & = \frac{1}{M_{\rm pl}^2} \int_{\chi_{\rm end}}^{\chi_{\rm ini}} d \chi \left( \frac{V(\chi)}{V'(\chi)} \right)
\,, \notag \\ 
\Delta_R^2& = \frac{V(\chi)}{24 \pi^2 M_{\rm pl}^4 \epsilon}
\,, 
\end{align}
with $M_{\rm pl}$ being the reduced Planck mass $\simeq 2.4 \times 10^{18}$ GeV. 
Since we work in an extremely tiny explicit-chiral (and scale) 
symmetry-breaking limit with $m_\pi \ll m_F ( \ll v_\chi)$ and the magnitude of the dilaton potential during the inflation (for $\chi \ll v_\chi$)
can be approximated by the vacuum energy $V_0$ as in the CW-SFI case, 
the slow-roll parameters are well approximately evaluated as 
\begin{align} 
\eta & = \frac{M_{\rm pl}^2}{V_0^{\rm LO}} \left( \frac{m_F}{v_\chi} \right)^2 \chi^2  
\Bigg[ - \frac{{72}}{\pi^2} \left( 1- 4 \pi^2 + 6 \pi^2 \ln \frac{\chi^2}{v_\chi^2} \right) 
\left( \frac{m_\pi}{v_\chi} \right)^2 
\notag \\ 
& 
+ \frac{384}{\pi^4} \left( 1 + \frac{3}{2} \ln \frac{\chi^2}{v_\chi^2}  \right) \left( \frac{m_F}{v_\chi} \right)^2
+ {\cal O}\left( \frac{m_\pi^4}{v_\chi^2 m_F^2} \right)
\Bigg] 
\,, \notag \\   
\epsilon & 
= \frac{M_{\rm pl}^2}{2 [V_0^{\rm LO}]^2} \left( \frac{m_F}{v_\chi} \right)^4 v_\chi^6  
\notag \\ 
& \times 
\Bigg[ \frac{{24}}{\pi^2} \left( \left\{1 - 12 \pi^2 \ln \frac{\chi^2}{v_\chi^2} \right\} \frac{\chi^3}{v_\chi^3} -1 \right) 
\left( \frac{m_\pi}{v_\chi} \right)^2  
\notag \\ 
& 
+ \left\{ \frac{192}{\pi^4} \frac{\chi^3}{v_\chi^3} \ln \frac{\chi^2}{v_\chi^2} \right\} 
\left( \frac{m_F}{v_\chi} \right)^2 
+ {\cal O}\left( \frac{m_\pi^4}{v_\chi^2 m_F^2} \right)
\Bigg]^2  
\,, 
\end{align}
with 
\begin{align} 
 V_0^{\rm LO} = \frac{24}{\pi^4} m_F^4
 \,. \label{V0LO}
\end{align}

The SFI with the extremely small chiral-scale breaking by the $m_\pi$ 
will give an overall scaling for $\epsilon/\eta$ 
with the small expansion factors as 
$ 
\frac{\epsilon}{\eta} \sim \left( \frac{m_\pi}{m_F} \right)^4 \left(
 \frac{v_\chi}{\chi} \right)^2 
 $. 
Hence the inflation would be ended by reaching $\eta =1$, as long as 
$\chi/v_\chi > (m_\pi/m_F)^2$, as in the CW-SFI case, 
which indeed turns out to happen as will be seen later.  
In that case (with $m_\pi \ll \chi \ll m_F \ll v_\chi$) 
, the $\eta$ and $\epsilon$ as well as 
the $\Delta_R^2$ and $N$ can further be approximated 
to be  
\begin{align} 
\eta & \simeq 24 \frac{M_{\rm pl}^2}{v_\chi^2} \frac{\chi^2}{v_\chi^2} \ln \frac{\chi^2}{v_\chi^2} 
\,, \notag \\ 
\epsilon & \simeq \frac{\pi^4}{{2}} \left( \frac{M_{\rm pl}}{v_\chi} \right)^2 
\left( \frac{m_\pi}{m_F} \right)^4 
\,, \notag \\ 
\Delta_R^2  & \simeq \frac{{2}}{\pi^{10}} 
\left( \frac{m_F}{v_\chi} \right)^4 \cdot \left( \frac{v_\chi}{M_{\rm pl}} \right)^6 
\left( \frac{m_F}{m_\pi} \right)^4 
\,, \notag \\ 
N & \simeq 
\frac{(\chi_{\rm end} - \chi_{\rm ini})}{\sqrt{2 \epsilon} M_{\rm pl}} 
\simeq 
\frac{(\chi_{\rm end} - \chi_{\rm ini}) v_\chi}{{6} \pi^2 M_{\rm pl}^2} 
\left( \frac{m_F}{m_\pi} \right)^2 
\,.  \label{approximations}
\end{align}

Note that 
 the gigantic suppression factor $\frac{2}{\pi^{10}} 
\left( \frac{m_F}{v_\chi} \right)^4$ for $\Delta_R^2$ in Eq.(\ref{approximations}) shows up, 
corresponding to an extremely tiny quartic coupling $\lambda_\chi$, 
realized by the walking nature $(m_\chi/v_\chi)^2 \ll 1$ as seen from 
the PCDC relation in Eq.(\ref{PCDC}) (also see Eq.(\ref{lambda-chi-mF})), 
which gets small enough to 
cancel the other factors coming from the small $\epsilon$, to easily 
achieve the right small amount of the observed $\Delta_R^2 \sim 10^{-9}$ at the pivot scale.  
Given the observed $\Delta_R^2$, the pion mass $m_\pi$ is actually 
written as a function of other potential parameters like 
\begin{align} 
 m_\pi^2 \simeq 
 \sqrt{\frac{2}{ \pi^{10} \Delta_R^2}} 
\left( \frac{m_F}{v_\chi} \right)^2
 \left( \frac{v_\chi}{M_{\rm pl}} \right)^3 m_F^2 
 \,. \label{mpi-DeltaR} 
\end{align}

Note also that the e-folding number $N$ in Eq.(\ref{approximations})
is set by the constant $\epsilon$, 
in contrast to the case of the CW-SFI where it is instead 
set by $\eta$ so that one would encounter the incompatibility between 
the $N$ and 
the spectral index $n_s \simeq 1+ 2 \eta$ in comparison with 
the observational values~\cite{Barenboim:2013wra,Takahashi:2013cxa}. 
As discussed in the literature~\cite{Iso:2014gka}, 
a small enough tadpole term (corresponding to the $C$-term at present) 
helps avoid this catastrophe, which will be more concretely demonstrated 
by the present walking inflationary model later on.

\section{Embedding into a dynamical scalegenesis} 

To more realistically analyze the present walking dilaton 
inflationary scenario, 
we need to consider couplings between the walking gauge 
sector and SM sector, so that we can access 
the reheating temperature $T_R$ and the e-folding number detected 
by the CMB photons at the pivot scale ($k=k_{\rm CMB}=0.05\,{\rm Mpc}^{-1}$) through the following relation~\cite{Aghanim:2018eyx}: 
\begin{align} 
 N_{\rm CMB} 
 \simeq 61 + \frac{2}{3}\ln \left(\frac{V_0^{1/4}}{10^{16} {\rm GeV}} \right) 
 + \frac{1}{3} \ln \left( \frac{T_R}{10^{16} {\rm GeV}}\right) 
 \,. \label{NCMB}
\end{align}
Aa a benchmark model, we shall try to 
embed the present scenario into 
a dynamical scalegenesis~\footnote{
The explicit-scale breaking $m_0$ term in Eq.(\ref{m0}) at the quantum level 
will not generate extra scale anomalies nor quadratic divergent 
contributions to the Higgs mass parameter $(m_H)$. 
Thereby one can keep 
the (almost) quantum scale invariance (up to the tiny $m_0$) 
up until the dimensional transmutation 
triggered in the hidden non-Abelian gauge sector,    
as long as 
the theory can be embedded into an asymptotic safety below or at  
the Planck scale~\cite{Gies:2003dp,Shaposhnikov:2008xi,Gies:2009sv,Braun:2010tt,Bazzocchi:2011vr,Wetterich:2011aa,Antipin:2013pya,Gies:2013pma,Tavares:2013dga,Abel:2013mya,Litim:2014uca,Litim:2015iea,Bond:2016dvk,Pelaggi:2017abg,Bond:2017wut,Barducci:2018ysr,Eichhorn:2018yfc,Abel:2018fls}, 
and the initial condition $m_H(M_{\rm pl})=0$ is 
realized by some over-Plankian nonperturbative dynamics~\cite{Shaposhnikov:2009pv,Wetterich:2016uxm,Eichhorn:2017als,Pawlowski:2018ixd,Wetterich:2019qzx}.}, 
in which 
the EWSB is triggered by what is called the bosonic seesaw mechanism~\cite{Calmet:2002rf,Kim:2005qb,Haba:2005jq,Antipin:2014qva,Haba:2015lka,Haba:2015qbz,Ishida:2016ogu,Ishida:2016fbp,Haba:2017wwn,Haba:2017quk,Ishida:2017ehu,Ishida:2019gri}~\footnote{
Embedding inflationary scenarios into dynamical 
scale generation mechanisms has extensively been addressed 
recently~\cite{Rubio:2017gty,Casas:2017wjh,Ferreira:2018qss,Ghilencea:2018thl,Barnaveli:2018dxo,Karam:2018mft,Kubo:2018kho,Shaposhnikov:2018nnm,Tang:2019uex,Ferreira:2019zzx,Tang:2019olx} in a context different from the present study. 
}. 
In that case, the hidden walking eight-flavor fermion fields 
$F^i$ ($i=1,\cdots, 8$) are vector-likely charged in part by 
the EW gauges. A possible charge assignment goes like 
\begin{align} 
 \Psi_{L/R} & = (F^1, F^2)^T_{L/R} & \sim (3, 1, 2, 1/2) 
 \notag \\ 
 \psi^{1,\cdots,6}_{L/R} & = F_{L/R}^{3,\cdots 8}  & \sim (3, 1, 1, 0)  
 \,,   \label{CA}
\end{align}
 under the SU(N) $\times$ SU(3)$_c \times$ SU(2)$_W \times$ U(1)$_Y$ symmetry.

The gauge invariance as well as the classical scale-invariance 
allows us to introduce a Yukawa coupling between 
the $F$-fermion fields and a Higgs doublet field $H$ as~\cite{Haba:2015qbz,Ishida:2016ogu,Ishida:2016fbp,Haba:2017wwn,Haba:2017quk,Ishida:2017ehu,Ishida:2019gri}  
\begin{align} 
{\cal L}_{y_H} = - \sum_{A=1}^6 y_H^{A} 
(\bar{\Psi}_L H \psi^A_R + \bar{\Psi}_R H \psi^A_L)   
+ {\rm h.c.}   
\,, \label{yH-Yukawa}
\end{align} 
(with the hidden-fermion parity invariance ensured by the underlying 
vectorlike gauge theory via Vafa-Witten theorem~\cite{Vafa:1983tf}), 
where the couplings $y_H$s are assumed to be small enough (to be consistent with 
the fitting later). 
After the chiral condensate (and confinement) develops, 
those $y_H$-Yukawa interactions 
generate a couple of mixings between the elementary $H$ doublet and composite Higgs 
doublets $\Theta^A \sim \bar{\psi}^A \Psi$, to give 
the negative mass square of the SM-like Higgs field (arising as the lowest 
mass eigenstate from the mass matrix), that is called the bosonic seesaw 
mechanism~\cite{Calmet:2002rf,Kim:2005qb,Haba:2005jq,Antipin:2014qva,Haba:2015lka,Haba:2015qbz,Ishida:2016ogu,Ishida:2016fbp,Haba:2017wwn,Haba:2017quk,Ishida:2017ehu,Ishida:2019gri}, as follows: 
\begin{align} 
m_H^2 \simeq  - \sum_{A=1}^6 [y_H^A]^2 m_F^2 \equiv - 6 y_H^2 m_F^2 
\,, 
\end{align} 
where the $y_H^A$ couplings have been assumed to be flavor universal ($y_H^A\equiv y_H$). 
Thus, with the quartic coupling for the $H$ at hand, 
the EWSB is dynamically achieved so that the 
SM-like Higgs acquires the EW vacuum expectation 
value ($v_{\rm EW} \simeq 246$ GeV) properly. 
Then, the $m_H$ scale is fixed by the 125 GeV Higgs mass as 
$m_H^2 = - m_{h(125)}^2/2 \simeq - (88\,{\rm GeV})^2$, so that the $y_H$ 
coupling is determined as a function of $m_F$ like 
\begin{align} 
 y_H^2 \simeq \left( \frac{36\,{\rm GeV}}{m_F} \right)^2 
\,. \label{cons-yH}
\end{align}

Another important point arising from the $y_H$ interactions is 
that Eq.(\ref{yH-Yukawa}) explicitly breaks the $F$-fermion chiral 
U(8)$_L \times$ U(8)$_R$ symmetry, even the vectorial part. 
The breaking effect of this non-vectorial type destabilizes   
the chiral manifold, yielding the instability for the pions (i.e. 
making the pions tachyonic)~\cite{Haba:2015qbz,Ishida:2016ogu,Ishida:2016fbp,Haba:2017wwn,Haba:2017quk,Ishida:2017ehu,Ishida:2019gri}. 
Although some pions charged by EW gauges get sizable enough 
masses on the order of ${\cal O}(\alpha_W m_F^2)$ to safely 
overcome this instability, other chargeless pions can be unstable.   
The tachyonic correction to those pion masses 
are evaluated by using the current algebra, 
as done in the literature~\cite{Ishida:2016ogu,Ishida:2016fbp}, to be 
\begin{align} 
 m_{y_H}^2 \simeq - (y_H v_{\rm EW}) \frac{\langle - \bar{F}F \rangle}{f_\pi^2}   
\,, \label{myH}
\end{align}
{where $\langle \bar{F}F \rangle$ denotes the chiral condensate per flavor}. 
This contribution has to be smaller than the pion-flavor universal $m_\pi$ term 
arising from the bare $m_0$ mass term for the $F$-fermions: 
$m_{y_H}^2 \ll m_\pi^2$~\footnote{
The sign ambiguity for the $y_H$ has been fixed to be positive 
by the definition of the chiral condensate 
$\langle \bar{F}F \rangle$ in Eq.(\ref{lad-cond}).  
}. 
The chiral condensate in the many-flavor walking gauge theory  
can be evaluated by a straightforward nonperturbative computation 
based on the Schwinger-Dyson equation analysis (in the ladder approximation)~\cite{Matsuzaki:2015sya}  
\begin{align} 
 \langle - \bar{F}F \rangle_{m_F} 
 \simeq \frac{8 N}{\pi^4} m_F^3 
 \,, \label{lad-cond}
\end{align}
in which the renormalization scale has been set at the scale $m_F$~\footnote{
Hence the $y_H$ coupling in Eq.(\ref{myH}) should also be interpreted as 
the one renormalized at the same scale $m_F$, so that 
the corresponding pion mass $m_{y_H}$ is independent of the renormalization 
scale, as it should be. 
Note also that the $y_H$ is not evolved by the renormalization below $m_F$ 
because of decoupling of the $F$-fermions, 
so the matching condition in Eq.(\ref{cons-yH}), which should be set at 
the 125 GeV Higgs mass scale, can be read as $y_H^2(m_F)=y_H^2(m_{h(125)})\simeq$ right hand side of Eq.(\ref{cons-yH}).}.   
Using Eqs.(\ref{cons-yH}) and (\ref{lad-cond}) together with Eq.(\ref{fpi}), 
the $m_{yH}$ is completely fixed to a constant: 
\begin{align} 
 m_{y_H}^2 \simeq - (169 \,{\rm GeV})^2 
 \,. 
\end{align}
Thus the walking pion (particularly for neutral pions) can safely be 
stabilized when 
\begin{align} 
 m_\pi \gg 169 \, {\rm GeV}  
 \,. \label{mpi-bound}
\end{align}

In the present inflationary scenario, 
the reheating temperature is actually determined by 
the walking pion decays to EW bosons: 
since the walking dilaton as the inflaton 
predominantly decays to walking pion pairs with 
the strong coupling, which will be much faster than 
the Hubble evolution of the Universe at that time, 
the rate of the SM particle production is controlled by 
the pion decays to the EW bosons coupled 
to the vector-likely charged $F-$fermion currents. 

The interactions between the walking pions and EW bosons, 
relevant to the pion decay processes,  
are completely determined by a covariantized 
Wess-Zumino-Witten term~\cite{Wess:1971yu,Witten:1983tx} (in a way analogously to the three flavor case discussed in~\cite{Ishida:2016ogu}): 
\begin{align} 
 {\cal L}_{\pi {\cal V}{\cal V}} 
 = - \frac{N}{4 \pi^2 f_\pi} 
 \epsilon^{\mu\nu\rho\sigma} 
 {\rm tr}[\partial_\mu {\cal V}_\nu \partial_\rho {\cal V}_\sigma \pi] 
 \,. 
\end{align} 
${\cal V}_\mu$ denotes the external gauge field of $8 \times 8$ matrix form, with 
the SU(2)$_W$ and U(1)$_Y$ gauge fields $(W_\mu^a, B_\mu)$ embedded following 
the charge assignment in Eq.(\ref{CA}), which is expressed as 
\begin{align} 
 {\cal V}_\mu 
& = \left( 
\begin{array}{cc} 
 [{\cal V}_\mu^{\rm EW}]_{2 \times 2} & 0_{2 \times 6} \\ 
0_{6 \times 2} & 0_{6 \times 6} 
\end{array}
 \right)
\,, \notag \\ 
 {\cal V}_\mu^{\rm EW} &= g_W W_\mu^a \tau^a + \frac{g_Y}{2}B_\mu\cdot 1_{2\times 2} 
 \,,  
\end{align}  
with the SU(2)$_W$ and U(1)$_Y$ gauge couplings, $g_W$ and $g_Y$, 
and normalized Pauli matrices $\tau^a$ with the normalization ${\rm tr}[\tau^a \tau^b]=\delta^{ab}/2$. The walking pion field $\pi$ is parametrized in a way similar to 
the ${\cal V}_\mu$ as 
\begin{align}
\pi=\left( 
\begin{array}{cc} 
 [\pi_{\Psi \Psi}]_{2 \times 2} & [\pi_{\Psi \psi}]_{2 \times 6} \\ 
 \left[\pi_{\psi \Psi}\right]_{6 \times 2} & [\pi_{\psi\psi}]_{6 \times 6} 
\end{array}
 \right)
 \,. \label{pis}
 \end{align}  
One can readily see that only the $\pi_{\Psi\Psi}$ component 
couples to the EW bosons. 
The EW charged pions in the $\pi_{\Psi\Psi}$ 
get large masses of ${\cal O}(g_W m_F)$, 
by the EW interaction, enough to close the decay channel 
of the walking dilaton into the pion pairs, as will be clarified later.  
Thereby, only the EW singlet one with the mass $m_\pi$, 
$\pi_{\Psi\Psi} \ni \pi_{\Psi\Psi}^0/2 \cdot 1_{2 \times 2}$, 
contributes to the determination of the reheating temperature~\footnote{
The possible mixing with the $\eta'$-like pseudoscalar would be small 
because of the large enough mass for the many-flavor walking $\eta'$ 
generated by the U(1) axial anomaly, $m_{\eta'} \sim \sqrt{N_F/N} m_F (\gg m_F)$, 
due to the anti-Veneziano limit~\cite{Matsuzaki:2015sya}. 
}.

The $\pi_{\Psi \Psi}^0$ total width is computed to be 
\begin{align} 
 \Gamma_{\pi_{\Psi \Psi}^0} = \frac{m_\pi^3}{16 \pi} \left( \frac{N}{4 \pi^2 f_\pi} \right)^2 
 \left[ \left( \frac{g_Y(m_\pi)}{2} \right)^2 + 3 \left(\frac{g_W(m_\pi)}{2} \right)^2 \right]
\,, \label{totwidth}
\end{align}
where we have taken the EW bosons to be massless because $m_\pi \gg m_{W, Z}$ as 
evident from Eq.(\ref{mpi-bound}). 
We also specified the renormalization 
scale for the EW gauge couplings at the $\pi_{\Psi \Psi}^0$ mass scale, 
which however tuns out to be the same as the ones evaluated at the $Z$ boson 
pole, $g_W(m_\pi)=g_W(m_Z)\simeq 0.42$ and $g_Y(m_\pi)=g_Y(m_Z) \simeq 0.13$~\footnote{
Those numbers have been estimated by using 
the electromagnetic couplings renormalized at the $Z$-boson mass scale  
($m_Z\simeq {91.2}$ GeV~\cite{Tanabashi:2018oca}), 
$\alpha(m_Z) = g_Y^2(m_Z) c_W^2/(4\pi) \simeq 1/128$~\cite{Tanabashi:2018oca} 
and the ($Z$-mass shell) Weinberg angle 
quantity $c_W^2 = m_W^2/m_Z^2 \simeq 0.778$. 
}, 
because possible 
renormalization corrections from EW-charged $F$-fermions, 
with the mass on the scale of 
$m_F$, necessarily get decoupled at the scale $m_\pi \ll m_F$. 
By equating the decay rate in Eq.(\ref{totwidth}) and the Hubble parameter 
with the radiation dominance, we determine the reheating temperature $T_R$ 
as follows: 
\begin{align} 
 T_R \simeq 0.23 \times \left(\frac{100}{g_*(T_R)} \right)^{1/2} \times 
 \sqrt{\Gamma_{\pi_{\Psi \Psi}^0} M_{\rm pl}}
\,, \label{TR}
\end{align}
with the effective degrees of freedom at $T_R$, $g_*(T_R)$, 
which is set to the SM value $= 106.75$~\cite{Tanabashi:2018oca}, as long as 
the walking dilaton and pion masses are larger than $T_R$, 
as in the present analysis to be clarified in the next section.

Thus the e-folding number for the CMB photons at the pivot scale 
($N_{\rm CMB}$) in Eq.(\ref{NCMB}) 
is evaluated as a function of $m_\pi$ and $m_F$, which has to be fitted to 
the theoretically predicted $N$ in Eq.(\ref{approximations}), so as for 
the present inflationary model to surely generate the 
currently observed CMB photons.

\section{Constraints and predictions} 

\begin{figure}
\begin{center}
\includegraphics[width=6cm]{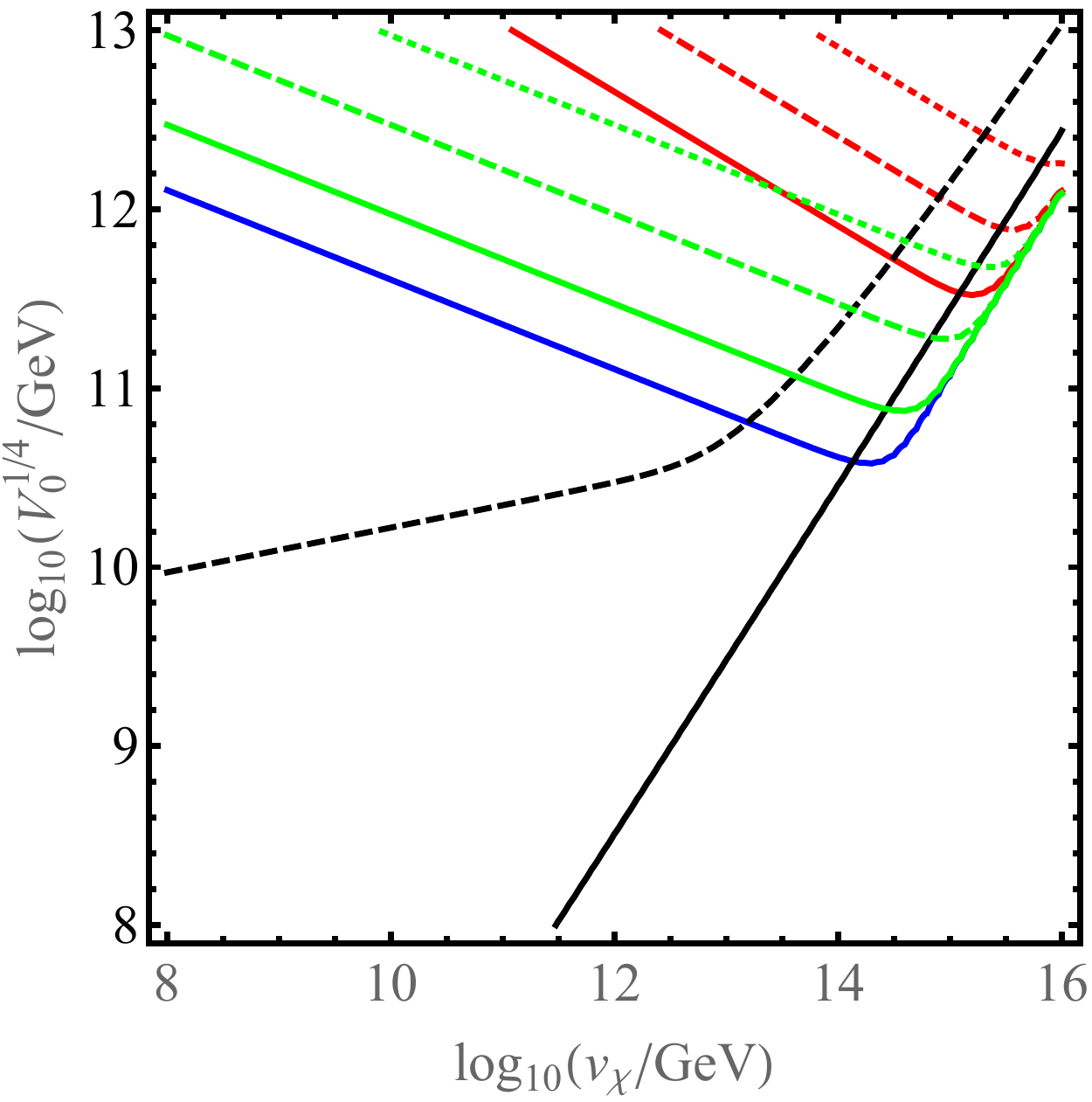}
\end{center}
\caption{
The exclusion plot on the plane ($v_\chi$, $V_0^{1/4}$).  
The black solid curve has been created by imposing 
$N=N_{\rm CMB}$ and the dashed one by $N=N_{30\, {\rm Gpc}}$~\cite{Tanabashi:2018oca}, 
below which the causally-connected patch, with 
the comoving distance ($d$) generated by the inflation, 
encompasses the observed Universe today, $d> 30$ Gpc. 
Hence the region sandwiched by those curves are allowed so that  
the required e-folding number can be realized. 
The pion stability is safe above the blue curve 
(see the main text for the detail). 
On solid, dashed, and dotted green curves, 
the reheating temperatures are 
$T_R = 10^2$, $10^3$, and $10^4$ GeV, respectively. 
On solid, dashed, and dotted red curves lines, 
the pion masses are $m_\pi = 10^3$, $10^4$, and $10^5$ GeV, respectively. 
}  
\label{summary-exclusion-plot-aboveTR100}
\end{figure}

Using the observed data for $\Delta_R^2$ and $n_s(=1 + 2 \eta)$, 
($\Delta_R^2 \simeq 2.137 \times 10^{-9}$ and $n_s \simeq 0.968$~\cite{Aghanim:2018eyx})  
with the initial stage of the inflation identified as the 
pivot scale, together with a couple of formulae derived in 
the previous sections, 
we now constrain the present walking-dilaton inflationary-scenarios  
embedded into a dynamical scalegenesis. 
Figure~\ref{summary-exclusion-plot-aboveTR100} 
shows the exclusion plot on the parameter space 
spanned by $v_\chi$ and $V^{1/4}_0$ (instead of $m_F$ with 
Eq.(\ref{V0LO}) taken into account).

We first see from the figure that 
the $v_\chi$ has to be greater than $10^{11}$ GeV in order to 
realize the right amount of the e-folding number for 
the CMB photons today by the inflation 
(i.e. the bound from $N$ in Eq.(\ref{approximations}) 
equals $N_{\rm CMB}$ in Eq.(\ref{NCMB})). 
Furthermore, the $v_\chi$ as well as the $V_0^{1/4} (\simeq m_F)$ 
are highly constrained by the pion instability set as in Eq.(\ref{mpi-bound}) 
to be $v_\chi \gtrsim 10^{14}$ GeV and 
$V_0^{1/4} (\simeq m_F) \gtrsim 3 \times 10^{10}$ GeV.

When the EW phase transition is assumed to happen in the thermal 
history of our Universe, the reheating temperature $T_R$ is necessary to be 
$\gtrsim 100$ GeV. In that case, 
the bounds on the $v_\chi$ and $V_0^{1/4}$ get 
shifted further upward, so that we would have 
\begin{align} 
v_\chi & \gtrsim  1.7 \times 10^{15}\,{\rm  GeV} 
\,, \notag \\ 
V_0^{1/4} (\simeq m_F) 
& \gtrsim 4.1 \times 10^{11}\, {\rm  GeV}
\,, \notag \\
& {\rm for} \qquad T_R  \gtrsim 10^2 \,{\rm GeV} 
\,. \label{outputs-1}
\end{align} 
Then the masses of the walking dilaton (in Eq.(\ref{Mchi})) and pion (in Eq.(\ref{mpi-DeltaR})) 
at those lower bounds 
are estimated to be  
\begin{align}
M_\chi 
(\simeq m_\chi) & \simeq 3.8 \times 10^8\,{\rm  GeV} 
\,, \notag \\  
m_\pi & \simeq 6.7 \times 10^4 \,{\rm GeV}
\,.  
\end{align} 
At this benchmark point, for other model parameters 
we have $\chi_{\rm ini} \simeq  6.7 \times10^9$ GeV, 
$\chi_{\rm end} \simeq 5.8 \times 10^{10}$ GeV, 
$\epsilon \simeq 5.1 \times 10^{-21}$, 
$N \simeq 51$, 
and $y_H \simeq 6.5 \times 10^{-11}$. 
One can easily see from those reference outputs 
that the approximated analytic formulae 
in Eq.(\ref{approximations}), derived for $m_\pi \ll \chi_{\rm ini, end} \ll m_F \ll v_\chi$,  
indeed work well, and surely reflects the large $N_F$ scaling 
(in the anti-Veneziano limit~\cite{Matsuzaki:2015sya}) intrinsic to 
the underlying many-flavor walking gauge theory,  
$v_\chi(\equiv f_\chi) \sim \sqrt{N_F} m_F \gg m_F$, and its associated pNG boson's nature 
$m_\chi, m_\pi \ll m_F$.

{In terms of the bare fermion mass $m_0$ in Eq.(\ref{m0}), 
one can also check the size of the explicit-chiral scale breaking regarding 
the $m_0$ to be extremely tiny: using the current algebra for the pion mass 
together with Eqs.(\ref{fpi}) and (\ref{lad-cond}), one gets $m_0/m_F 
= m_\pi^2 f_\pi^2/(2 \langle - \bar{F}F \rangle m_F) 
\simeq \pi^2/64 (m_\pi/m_F)^2$, 
which is $\simeq 4.1 \times 10^{-15}$ at the benchmark point above}. 
This tiny $m_0$ would be physical to be probed by detecting light walking pions with 
the mass $\gtrsim 10^3$ GeV.

In addition, we observe that the non-Gausiannity is small enough 
to be consistent with the latest results by Planck satellite~\cite{Akrami:2019izv}. 
The non-Gaussianity can be evaluated as $f_{\rm NL} = (5/12) (n_s + f(k) n_t)$ 
where $n_s (\simeq 0.96)$ and $n_t$ are the spectral indices for scalar and tensor modes, 
and $f(k)$ is determined by the shape of triangle which has a range of values $0 \leq f \leq 5/6$~\cite{Maldacena:2002vr}. 
Since $n_t$ and the scalar to tensor ratio $r$ have a relation: $r+8n_t=0$, 
the second term in the parenthesis gives a negative contribution to the non-Ganssianity. 
Namely, the maximal value of $f_{\rm NL}$ can be obtained to be 
roughly $0.4$, which is within the current limit: $f_{\rm NL} = -0.9 \pm 5.1$ 
(at the $68\%$ confidence level)~\cite{Akrami:2019izv}. 
Therefore, the non-Gaussianity is also consistent with the observation.

\section{Summary and discussion}

In summary, we have proposed an inflationary scenario of the CW-SFI type, 
dynamically arising from a large $N_F$ walking gauge theory, what we called 
the hidden walking gauge theory. 
The inflation is played by the walking dilaton and 
the inflaton potential parameters 
are highly constrained by the walking nature. 
We have evaluated the potential parameters 
using outputs from our best knowledge based on  
straightforward nonperturbative analyses. 
We showed that due to the intrinsic feature of the 
large $N_F$ walking dynamics, called the anti-Veneziano-large $N_F$ 
scaling, the desired tiny inflaton quartic coupling can 
naturally be realized, and the inflaton coupled to 
the walking pions can survive the 
cosmological, astrophysical constraints, 
from which other models of the CW-SFI generically suffer.

For the inflationary scenario to be realistic, 
as a reference model, the hidden walking dynamics 
was vector-likely 
gauged in part by the EW charges and was coupled to a Higgs doublet in 
a scale-invariant way, so that  
the EW symmetry breaking is triggered by the bosonic seesaw mechanism. 
This benchmark model has been fitted 
with all the cosmological and astrophysical constraints presently placed by 
Planck 2018 data and 
theoretical requirements on the walking dilaton and chiral pion physics.   
We found that the present inflationary scenario is highly constrained 
to give a stringent bound on the fermion dynamical mass 
scale to be $> 10^{11}$ GeV, and  
the reheating temperature ($T_R$), which is fixed 
by chiral pion decays to EW 
gauge bosons including photons, 
to be $\gtrsim 10^2$ GeV. 
It also turned out that the masses of walking dilaton and pions  
are constrained by several theoretical and astrophysical limits  
to be $\sim 10^8$ GeV and $\sim 10^5$ GeV, respectively.


In closing, we shall give comments on issues to be left in the future works.

{
The present walking inflationary scenario works 
for a so large dilaton decay constant $v_\chi(\equiv f_\chi)$ 
compared to the pion decay constant $f_\pi$, $v_\chi \gtrsim 10^4 f_\pi$. 
The current lattice simulations~\cite{Aoki:2016wnc} have tried to 
observe those quantities with the 
current fermion mass $m_0/m_F\simeq 0.18 - 0.45$, to give a preliminary 
result on the dilaton decay constant $f_\chi(\equiv v_\chi) \simeq 3.7 f_\pi$ 
based on the lowest-order formula in the dilaton-chiral perturbation theory~\cite{Matsuzaki:2013eva}, 
by simply fitting the dilaton mass formula as in Eq.(\ref{Mchi}) with 
the data on the slope with respect to the leading order $m_\pi^2$ dependence 
on the dilaton mass ($d M_\chi^2/d m_\pi^2$).  
(This has been consistent with a direct measurement of the pole residue 
of the scalar current correlator with a simple-minded linear fit assumed 
in the same range of the current fermion mass.). 
However, 
this result cannot simply be compared with our values at present: 
for our reference-current fermion-mass value $m_0/m_F \sim 10^{-15}$, 
the next-to leading-order chiral-logarithmic correction 
would be significant for the 
estimates on the dilaton mass, and decay constant cannot be 
measured just by the slope $d M_\chi^2/d m_\pi^2$,  
as explicitly demonstrated in~\cite{Matsuzaki:2013eva}. 
Future upgraded lattice setups, by which 
the chiral log corrections are visible, could check if our scenario is indeed 
viable on the ground of the realistic nonperturbative dynamics}.

The phenomenological consequence for the 
present inflationary scenario, distinguishable from other 
inflation models in the huge ballpark, would be derived from 
the presence of walking pions with mass $\gtrsim 10^3$ GeV 
(if an EW phase transition happens by supercooling like in the scenario 
discussed in the literature~\cite{Iso:2017uuu}). 
Such a sub TeV-walking pions can have several predictions for terrestrial and satellite experiments. 
At current or future hadron collider experiments, 
the walking pions are potentially produced via photon fusion process~\cite{Ishida:2016ogu}. 
The detail analyses for these experimental signatures are to be 
discussed in future publications.

Moreover, actually we can have rich amount (35 kinds) of dark matter candidates 
(corresponding to the $\pi_{\psi\psi}$ states in Eq.(\ref{pis}), 
except the heavy walking $\eta'$ decaying to EW bosons just like 
the $\pi_{\Psi\Psi}^0$). 
Hence the thermal history and testability at underground and satellite experiments 
for those dark matter particles are deserved to be addressed as well.

{
As noted in the Introduction, 
the initial condition for the inflaton has been set to be away (to the left side) 
enough from the vacuum in the potential, by assuming some trapping mechanism 
to work there, as discussed in the literature~\cite{Iso:2015wsf} 
for the CW-SFI scenario. 
Actually, it would not be so plausible to apply the existing trapping mechanism: 
first, in our benchmark scenarios the inflaton gets coupled to the SM-like Higgs via 
a Higgs portal coupling as noted in~\cite{Ishida:2019gri}, which would be 
crucial for the trapping to work~\cite{Iso:2015wsf}. 
Note the size of the portal coupling ($\lambda_{\rm mix}$) 
turns out to be extremely smaller 
than the standard-Higgs quartic coupling of ${\cal O}(10^{-1})$: $\lambda_{\rm mix} 
= m_H^2/v_\chi^2 \lesssim 10^{-26}$ (see Eq.(\ref{outputs-1})), hence the Higgs parametric 
resonance may not work well for trapping the $\chi$ at around the origin of the potential 
as argued in~\cite{Iso:2015wsf}. 
However, in contrast to the literature having arguments based on the potential 
for elementary scalar fields, 
another trapping possibility other than coupling to the SM Higgs 
might be present in the case of the composite inflaton we have employed, 
which could be 
closely tied with the underlying nonperturbative gauge dynamics. 
Though being beyond the current scope, 
this issue will involve some generic features and expected developments on 
the particle production mechanism, so it would be worth pursuing in the future. 
}

\section*{Acknowledgements} 
We are grateful to Satoshi Iso
and Kazunori Kohri for useful comments. 
We are also grateful to Seishi Enomoto for valuable discussions. 
H.I. thanks for the hospitality of Center for Theoretical Physics and College of Physics, 
Jilin University where the present work has been partially done. 
This work was supported in part by the National Science Foundation of China (NSFC) under Grant No. 11747308 and 11975108, 
and the Seeds Funding of Jilin University (S.M.).  
The work of H.I. was partially supported by JSPS KAKENHI Grant Number 18H03708.





\end{document}